# Post glacial rebounds measure the viscosity of the lithosphere

Jozsef Garai

Department of Earth Sciences
Florida International University
University Park, PC-344
Miami, FL 33199
Ph: 305-348-3445
Fax: 305-348-3070
E-mail: jozsef.garai@fiu.edu

**Abstract:** The observed higher uplift rates before the end of deglaciation requires the existence of a low viscosity channel or layer. The uplifts observed after the end of deglaciation does not show any contribution from this low viscosity channel and a homogeneous viscosity model fits very well to the observed uplift. Most of the researchers therefore prefer the homogeneous model and suggest that the higher uplift rate before the end of deglaciation is the result of elastic contamination. It has been shown that the elastic deformation of the lithosphere is far too small to be responsible for the observed extra uplift; therefore, the homogeneous viscosity model should be discredited.

The homogeneous viscosity of the postglacial period and the high uplift rate of the late glacial period can be explained with a model which has an upper layer determining the homogeneous viscosity and the layer below it which has a low viscosity. The contribution to the uplift of this low viscosity layer is indistinguishable from an instantaneous uplift.

Current GPS deformation measurements indicating $10^{21}$ Pas average viscosity for the lithosphere, which is consistent with this model. The viscosity of the low velocity zone should be between $10^{11}$ Pas and $10^{18}$ Pas. The lower constrain on the viscosity of the low velocity zone can be deduced from seismic observations. This viscosity model is consistent with post glacial rebounds and further supported by the lack of the dynamic topography.

## 1. Introduction

In 1865, T.F. Jamieson recognized and reported evidence that the uplifted beaches of Fennoscandia were the result of the removal of the ice sheet. Similar movements of the North American crust were reported by Bell in 1896 and by Gilbert in 1897. They concluded that warmer weather melted a great amount of ice on the earth's surface, causing huge mass redistribution. The lithosphere responded to this mass redistribution with deformation, which can be characterized by time-altitude graphs, where the altitude is compared to the eustatic sea-level. Previous investigations (Andrews, 1968; Farrand, 1962; Schofield, 1964) identified the following characteristic features of the uplift curves:

- there was a rapid, 5-10 (or even higher) cm/year uplift, which started in each area soon after the deglaciation;
- during this fast initial uplift period, the uplift developed independently from the general uplift of the overall area;
- the rate of uplift sharply decreased at a given locality immediately following the deglaciation of that locality;
- after the completion of the deglaciation, the subsequent uplift developed in harmony with the general uplift and the rate of the uplift can be characterized with a simple exponential decrease.

These identified features of the uplifts are independent of the locality, although the curves are displaced in time, and this displacement correlates with the time of deglaciation of each locality. The time of deglaciation divides the uplifts into two distinguishable parts. In the first part the uplift rate is fast and independent from the general uplift, while in the second part the uplift rate is much slower, and can be characterized by an exponential decrease. Based on these significantly different uplift characteristics Walcott (1972) even suggested different names for these periods. For the North American Uplift he suggested that the uplift for the period 18,000-6,500 B.P. (before present) should be called late glacial, while the uplift from 6,500 to the present shall be named as postglacial.

The higher uplift rate in the late glacial period requires a low viscosity layer or channel with a viscosity of $10^{18} - 10^{19}$ Pas (Anderson, 1967; Artushkov, 1971; Fjeldskaar, 1994; Walcott, 1973). The presence of a low viscosity layer is supported by geological and seismological observations, which favors a very week asthenosphere, and by the observed relative submergence around the regions of the postglacial rebounds (Walcott, 1972). The uplift of the postglacial period can be explained with a homogenous average viscosity of $10^{21}$ Pas (Haskell, 1935; Mitrovica, 1996; Peltier, 1976). The excellence fit between the homogeneous viscosity models and the observed uplift curves excludes the possibility of a low viscosity zone. The proposed viscosity models are either able to explain the late glacial or postglacial period but not both. The debate between the two different interpretations of the postglacial rebounds has not been settled (Fjeldskaar, 2000; Lambeck, 2000) and comprehensive model has not been proposed.

## 2. The uncertainty of the viscosity

The low viscosity layer and the homogeneous viscosity models are mutually exclusive; therefore, if either of these models is correct then the other one has to be discredited. Researcher promoting homogeneous viscosity model claim that the extremely good fit between the model and the uplift, observed during the postglacial period, excludes the possibility of a low viscosity zone.

Any two independent set of data of the time-altitude observations allows one to calculate the viscosity for the time period elapsed. The uncertainty of these calculated viscosities gives information on the homogeneity of the viscosity. If the uncertainty ($2\sigma$) is high then the homogeneous viscosity model is not supported, while a low uncertainty would support the model. The viscosities were calculated form the best documented uplift observed at the mouth of the Angerman river in Sweden (Linden, 1938). There is a general agreement (Mitrovica, 1997) that the uplift process can be characterized by a simple exponential decay process for the region in which isostatic equilibrium was reached. The Angerman river area satisfies this condition, since it is located in the middle of the Fennoscandian uplift. One of the simplest formula, developed for half infinite homogeneous viscous substratum (Heiskanen, 1958; Meinesz, 1937) is

$$h_n = h_o e^{\frac{-t_n}{\tau_R}} \tag{1}$$

where, $h_o$ is the total amount of uplift that will eventually be achieved, $h_n$ is the uplift remaining at the time of $t_n$, where $t_n$ is time in years since the application of the load, $\tau_R$ is the relaxation time in years required for the deviation from isostasy to be reduced to 1/e of its initial value, where e is the base of the natural logarithm. The remaining uplift at the time of $t_n$ is

$$h_n = h_{n-observed} + h_{remaining} \tag{2}$$

where $h_{n-observed}$ is the observed uplift from the time $t_n$ until today and $h_{remaining}$ is the uplift remaining from the present time until the full completion of the uplift. Substituting equation two into one, the two unknowns, relaxation time, and remaining uplift can be calculated from any two independent sets of observations. Using the data of Angerman river the average of the remaining uplift is 46 m, while the average of the relaxation time is 5319 years for the period between 7944 and 87 years before present (Fig. 1). These values are in good agreement with previous studies (Mitrovica, 1996).

Assuming that the viscosity is Newtonian and that the mantle is uniform in a half-infinite space under the uplift area, the viscosity [$\eta$] can be calculated as:

$$\eta = \frac{\tau_R \rho g \lambda}{4\pi} \tag{3}$$

where $\rho$ is the density of the mantle, g is the gravitational acceleration, and $\lambda$ is the wavelength of the uplift. For the Fennoscandian uplift the following values were used: $\tau_R$ = 5319 years, $\rho$ = 3,370 kgm$^{-3}$, g = 10 ms$^{-2}$, $\lambda$ = 2,000 km.
The calculated average viscosity under the Fennoscandian area is 0.88x10$^{21}$ Pas (Fig 1). The uncertainty of the calculated viscosity is 0.21x10$^{21}$ Pas. This relatively small uncertainty indicates that a homogeneous model can fully explain the observed uplift for the postglacial period from 7944 to 87 BP years and that the possible contribution of a low viscosity layer to the uplift can be excluded. This also means that the low viscosity models should be discredited.

## 3. The size of the elastic rebound

The homogeneous viscosity model consistent with the uplift observed in the postglacial period but can not explain the observed fast uplift during the late glacial period. Researchers promoting these models interpret the higher uplift rate during the late glacial period as the result of the elastic deformation of the lithosphere (Mitrovica, 1996). The size of the elastic deformation of the lithosphere will be considered in details.

Assuming a homogeneous viscosity and using previously determined relaxation time the total uplift for the entire time period can be calculated. Using the most conservative approach, assuming that the full load was released immediately at the beginning of the deglaciation, the total uplift is 255.2 m. The least conservative approach would be to assume that the ice was melted at the end of the late glacial period which would result an uplift of 203.7m. The difference between these calculated uplifts and the observed maximum uplift is 100-150m. Thus the expected elastic deformation for the Fennoscandian region should be between 100 m. and 150 m (Fig. 2).

When a load from the surface is released the lithosphere undergoes two kinds of elastic deformations, thickening and unbending. From these two kinds of deformation only the thickening can occur instantaneously, since the unbending is controlled by the plastic deformation of the mantle (Fig. 3). In calculating the elastic deformation, the thickening of a unit in the vertical direction $\varepsilon_z$ is

$$\varepsilon_z = \frac{\frac{(1+\nu)(1-2\nu)}{1-\nu}\sigma_z}{E} \tag{4}$$

The total thickening [S] is

$$S = z\,\varepsilon_z \tag{5}$$

Where, $\sigma_z$ is the pressure release caused by the melted ice, z is the thickness of the lithosphere, E is the Young's modulus, and $\nu$ is the Poisson's ratio. For simplicity, the load is infinite in the horizontal extent, and the sides of the loaded column are restrained by hydrostatic pressure, in which case the elongation in the horizontal plane will be negligible. If the lithosphere is in isostatic equilibrium then the stress release on the surface will not have an affect on the stress of the mantle; therefore, the thickness of the thickening layer is equivalent with the thickness of the lithosphere. The stress in the mantle will decrease in accordance with the uplift of the lithosphere. For the calculation, the following values were used: $\sigma_z$ = 20 MPa, equivalent to about 2 km ice sheet; z = 75 km; E = $8.35 \times 10^{10}$ N/m$^2$; $\nu$ = 0.25.

The resulting thickening of the lithosphere, which should be observed on the surface from the stress release caused by the melting of 2 km thick ice sheet, is about 18 m. This value is far less than the observed access deformation of 100-150 m. The elastic deformation of the lithosphere can not be responsible for the extra uplift of the late glacial period; therefore, the homogeneous viscosity models should be discredited.

## 4. Proposed viscosity model

The only solution able to satisfy both the late glacial and postglacial uplifts is a viscosity model in which the viscosity of the upper later is $10^{21}$ Pas while the viscosity below this layer is significantly lower. This proposed viscosity for the upper layer is consistent with current GPS deformation measurements which indicate that the lithosphere has an average viscosity of $10^{21}$ Pas (Flesch et al., 2000).

The low viscosity layer under the lithosphere allows the unbending of the lithosphere very quickly. This fast deformation in geological scale is practically indistinguishable from elastic deformation. Seismic observations can constrain the lower limit of the viscosity of the asthenosphere. Using one second for the Maxwell relaxation time and 100 GPa for the elastic modulus, the lowest viscosity value for the low velocity zone is $10^{11}$ Pas.

A model with a low viscosity layer $(10^{18}\,\text{Pas} \geq \eta \geq 10^{11}\,\text{Pas})$, and with a viscosity of $10^{21}$ Pas for the lithosphere, can fully explain the observed uplift of postglacial rebounds at any sites. Additionally, the very low viscosity layer is consistent with the lack of the dynamic topography. Based on the currently accepted viscosity models, the mantle flow in the asthenosphere should maintain a much larger dynamic topography, which is not observed. This is a major unexplained problem in geophysics (Wheeler, 2000). If the viscosity directly below the lithosphere is decreased by several orders of magnitude the amplitude of the dynamic topography on the surface of the Earth is reduced, but only by

~20% (Lithgow-Bertolloni, 1997). A more dramatic decrease in the viscosity of the low viscosity layer might explain the complete lack of the dynamic topography.

## 5. The rigidity of the lithosphere

There is one argument that might be raised against the very low viscosity of the asthenosphere. The rigidity of the lithosphere based on previous investigations seems to be too small (Brochie, 1969) to explain the observed flexure (Morner, 1980). There are uncertainties in these previous investigations, like the distribution of the ice load is unknown. If the ice load had a close to a parabolic distribution (Brochie and Silvester, 1969) instead of a uniform one then the crustal deflection becomes very close to the observed uplift of the Fennoscandian area.

Rupture of the crust can also reduce the flexure of the lithosphere by reducing the wavelength of the uplift. Whole crustal rupturing, related to the postglacial rebound, has been reported recently (Arvidson, 1996).

Isostatic pressure from the bottom of the lithosphere compensates the ice load. If the entire area is in isostatic equilibrium and the lithosphere supports the unbalanced load within its rigidity limit then the elastic deformation of the lithosphere should be reduced greatly.

Superimposing these previous affects, parabolic load distribution, rupture of the crust, and isostatic equilibrium for the entire uplift area, can result a lithospheric flexure which is consistent with the observed ones.

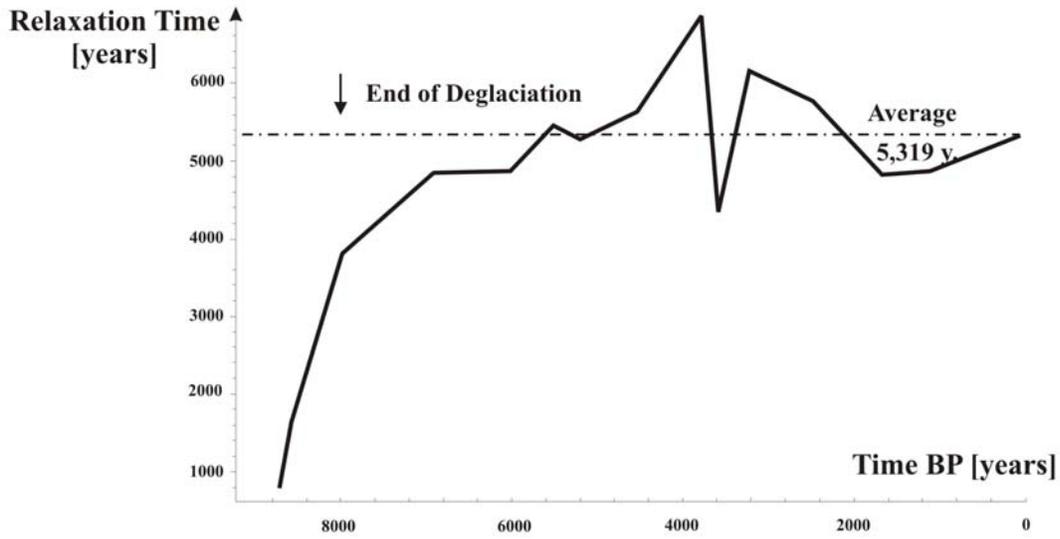

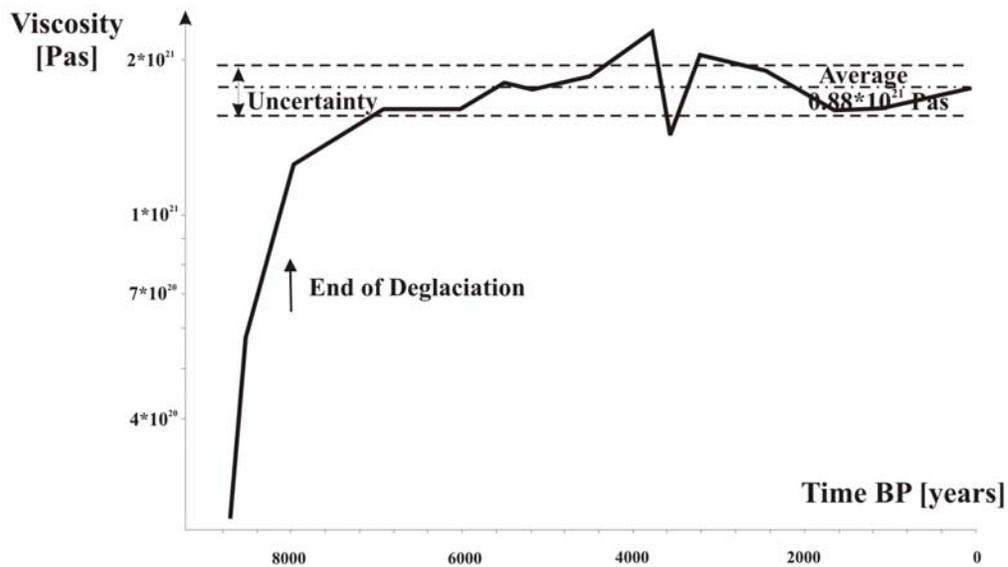

Fig. 1

Fig. 1 Calculated relaxation times and viscosities for the Fennoscandian uplift. The values were determined from every two consecutive sets of independent data, observed at the Angerman river.
    a./ relaxation times
    b./ viscosities

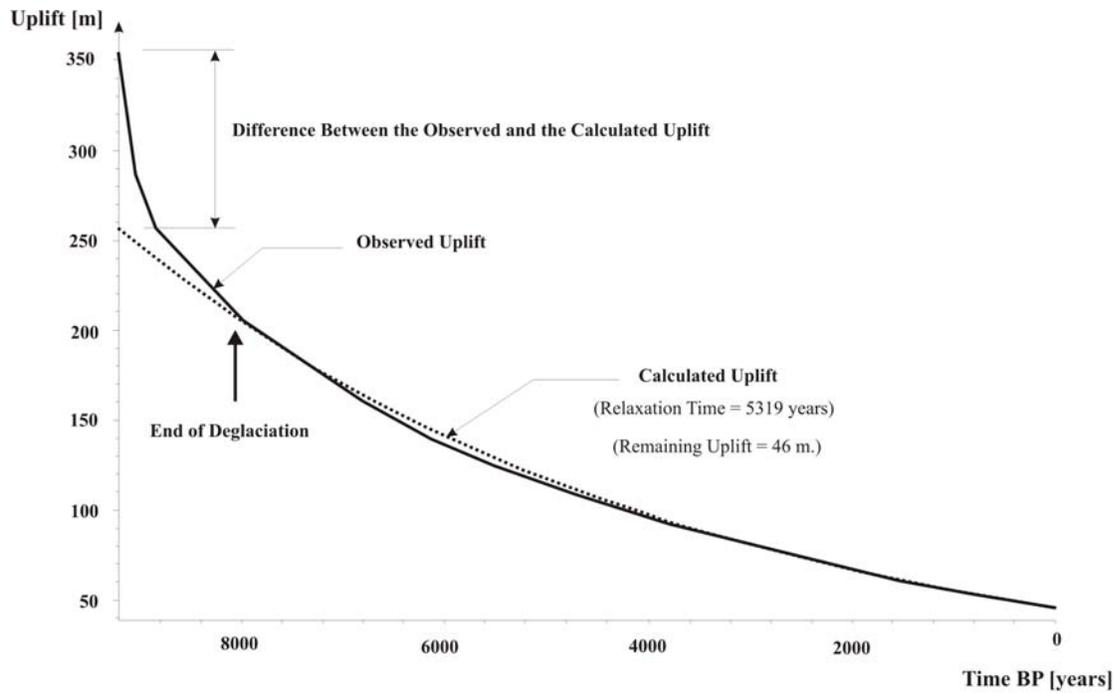

Fig. 2 Observed and calculated uplifts. For the calculation it was assumed that the full load was released immediately at the beginning of the deglaciation. The difference between the calculated uplift and the observed maximum uplift should be equivalent with the elastic deformation for the Scandinavian region.

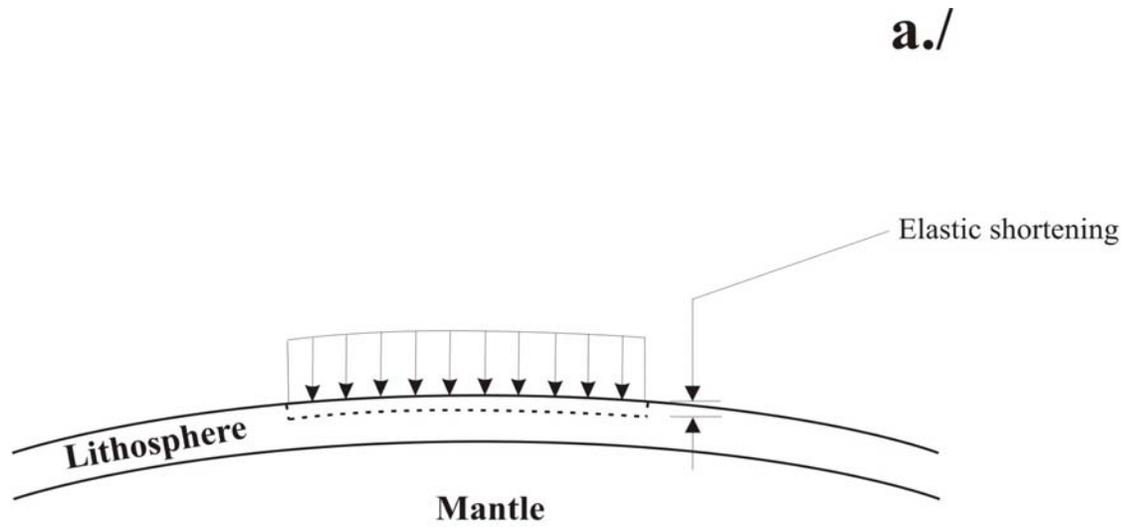

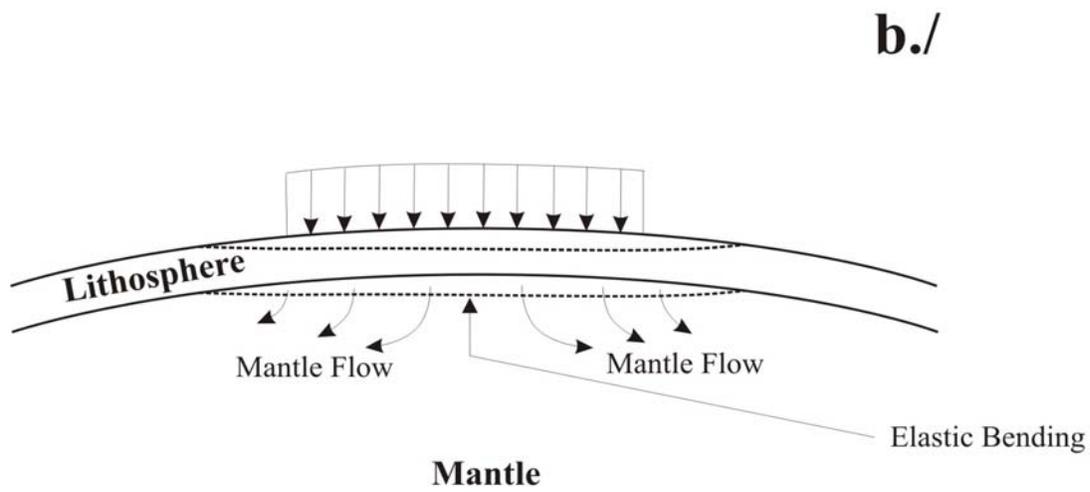

Fig. 3

Fig. 3 Elastic deformations of the lithosphere.
    a./ Elastic shortening can occur freely.
    b./ Elastic bending is restricted by the mantle flow.